# Graph Isomorphism for Graph Classes Characterized by two Forbidden Induced Subgraphs

Stefan Kratsch
Utrecht University, the Netherlands
s.kratsch@uu.nl

Pascal Schweitzer
The Australian National University
pascal.schweitzer@anu.edu.au

August 8, 2018

**Abstract**

We study the complexity of the Graph Isomorphism problem on graph classes that are characterized by a finite number of forbidden induced subgraphs, focusing mostly on the case of two forbidden subgraphs. We show hardness results and develop techniques for the structural analysis of such graph classes, which applied to the case of two forbidden subgraphs give the following results: A dichotomy into isomorphism complete and polynomial-time solvable graph classes for all but finitely many cases, whenever neither of the forbidden graphs is a clique, a pan, or a complement of these graphs. Further reducing the remaining open cases we show that (with respect to graph isomorphism) forbidding a pan is equivalent to forbidding a clique of size three.

## 1 Introduction

Given two graphs $G_1$ and $G_2$, the Graph Isomorphism problem (GI) asks whether there exists a bijection from the vertices of $G_1$ to the vertices of $G_2$ that preserves adjacency. This paper studies the complexity of GI on graph classes that are characterized by a finite number of forbidden induced subgraphs, focusing mostly on the case of two forbidden subgraphs. For a set of graphs $\{H_1, \ldots, H_k\}$ we let $(H_1, \ldots, H_k)$-free denote the class of graphs $G$ that do not contain any $H_i$ as an induced subgraph.

As a first example, consider the class of graphs containing neither a clique $K_s$ on $s$ vertices, nor an independent set $I_t$ on $t$ vertices. Ramsey's Theorem [18] states that the number of vertices in such graphs is bounded by a function $f(s,t)$. Thus the classes $(K_s, I_t)$-free are finite and Graph Isomorphism is trivial on them. All other combinations of two forbidden subgraphs give graph classes of infinite size, since they contain infinitely many cliques or independent sets.

As a second example, consider the graphs containing no clique $K_s$ on $s$ vertices and no star $K_{1,t}$ (i.e., an independent set of size $t$ with added universal vertex adjacent to every other vertex). On the one hand this class contains all graphs of maximum degree less than $\min\{s-1,t\}$, on the other hand, all graphs in $(K_s, K_{1,t})$-free have bounded degree: Indeed, if the degree of a vertex is sufficiently large, its neighborhood must contain a clique of size $s$ or an independent set of size $t$ by Ramsey's Theorem [18], leading to one of the two forbidden subgraphs. Thus, using Luks' algorithm [15] that solves Graph Isomorphism on graphs of bounded degree in polynomial time, isomorphism of $(K_s, K_{1,t})$-free graphs can also be decided in polynomial time.

To systematically study Graph Isomorphism on graph classes characterized by forbidden subgraphs, we ask: *Given a set of graphs $\{H_1, \ldots, H_k\}$, what is the complexity of Graph Isomorphism on the class of $(H_1, \ldots, H_k)$-free graphs?*

**Related work.** The Graph Isomorphism problem is contained in the complexity class NP, since the adjacency preserving bijection (the isomorphism) can be checked in polynomial time. No polynomial-time algorithm is known and it is known that Graph Isomorphism is not NP-complete unless the polynomial hierarchy collapses [5]. More strongly, Graph Isomorphism is in the low hierarchy [20]. This has led to the definition of the complexity class of problems polynomially equivalent to Graph Isomorphism, the so-called GI-complete problems. There is a vast literature on the Graph Isomorphism Problem; for a general overview see [21] or [10], for results on its parameterized complexity see [12].

A question analogous to ours, asking about Graph Isomorphism on any class of $(H_1, \ldots, H_k)$-minor-free graphs, is answered completely by the fact that Graph Isomorphism is polynomially solvable on any non-trivial minor closed class [17]. Recently, the corresponding statement for topological minor free classes was also shown [8]. For the less restrictive family of *hereditary* classes, only closed under vertex deletion (i.e., classes $\mathcal{H}$-free for a possibly infinite set of graphs $\mathcal{H}$), both GI-complete and tractable cases are known: Graph Isomorphism is GI-complete on split graphs, comparability graphs, and strongly chordal graphs [22]. Graph Isomorphism is known to be polynomially solvable for circle graphs and circular-arc graphs [9], interval graphs [2, 14], distance hereditary graphs [16], and graphs of bounded degree [15]. For various subclasses of these polynomially solvable cases results with finer complexity analysis are available, but of course the polynomial-time solvability for these subclasses follows already from polynomial-time solvability of the mentioned larger classes. Further results, in particular on GI-completeness, can be found in [4].

Concerning our question, for one forbidden subgraph, the answer, given by Colbourn and Colbourn, can be found in a paper by Booth and Colbourn [4]: If the forbidden induced subgraph $H_1$ is an induced subgraph of the path $P_4$ on four vertices, denoted by $H_1 \leq P_4$, then Graph Isomorphism is polynomial on $H_1$-free graphs, otherwise it is GI-complete.

Apart from the isomorphism problem, other studies aiming at dichotomy results for algorithmic problems on graph classes characterized by two forbidden subgraphs consider the chromatic number [11] and dominating sets [13].

**Main result.** Let a graph be *basic* if it is an independent set, a clique, a $P_3 \dot{\cup} K_1$, or the complement of a $P_3 \dot{\cup} K_1$ (also called pan). If neither $H_1$ nor $H_2$ is basic, then we obtain a classification of $(H_1, H_2)$-free classes into polynomial and GI-complete cases, for all but a small finite number of classes. Theorem 1 justifies the terminology basic by showing that in our context forbidding a basic graph is equivalent to forbidding a complete graph. However, the case of forbidding a clique (alongside a second graph) appears to be structurally different and for a complete classification further new techniques are required.

**Technical contribution.** Our main technical contribution lies in establishing tractability of Graph Isomorphism on four types of $(H_1, H_2)$-free classes (Theorem 4 in Section 4): A structural analysis of the classes enables reductions to the polynomially-solvable case of bounded color valence [1]. This reduction appears necessary since the polynomially-solvable classes of Theorem 4 encompass all classes of graphs of bounded degree, and for these Luks' group-theoretic approach [15] (implicit in [1]) is the only known polynomial-time technique. At the core of the proof of Theorem 4 lie individualization-refinement techniques and recursive structural analysis to allow for a reduction to the bounded color valence case.

However, to put these results in context and obtain the mentioned classification, we have to refine several known results for GI-completeness on bipartite, split, and line graphs (Section 3). In particular, we arrive at a set of four graph properties, which we call *split conditions*, such that Graph Isomorphism remains complete on any class $(H_1, H_2)$-free unless each property is true for at least one of the two forbidden subgraphs.

Based on this characterization we can state our results in more detail: If on the one hand

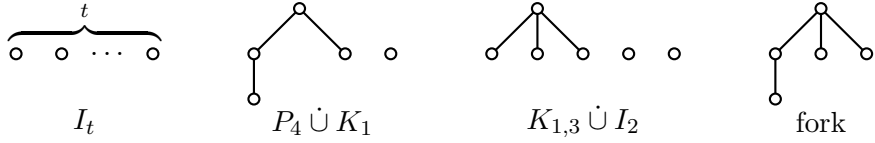

$I_t$     $P_4 \mathbin{\dot\cup} K_1$     $K_{1,3} \mathbin{\dot\cup} I_2$     fork

Figure 1: The independent sets, the paths of length 3 with independent vertex, the claw with two independent vertices, and the fork, obtained by subdividing an edge in a claw.

neither of the two forbidden subgraphs $H_1$ and $H_2$ exhibits all four split conditions, then we have a dichotomy of GI on $(H_1, H_2)$-free classes into polynomial and GI-complete classes; the polynomial cases are due to Theorem 4 (see Section 4) as well as tractability on cographs (i.e., $P_4$-free graphs) [7], GI-completeness follows by using both known results as well as our strengthened reductions (see Section 3). Suppose on the other hand $H_1$ and $H_2$ are both not basic and $H_1$ simultaneously fulfills all four split conditions, then our hardness and tractability results resolve all but a finite number of cases (i.e., each case is one concrete class $(H_1, H_2)$-free), as showed in Theorem 6 (see Section 5). For these cases Figure 1 shows the relevant maximal graphs that adhere to all four split conditions.

## 2 Preliminaries

We write $H \leq G$ if the graph $G$ contains a graph $H$ as an induced subgraph. A graph $G$ is $H$-free if $H \not\leq G$. It is $(H_1, \ldots, H_k)$-free, if it is $H_i$-free for all $i$. A graph class $\mathcal{C}$ is $H$-free (resp. $(H_1, \ldots, H_k)$-free) if this is true for all $G \in \mathcal{C}$. A graph class $\mathcal{C}$ is *hereditary* if it is closed under taking induced subgraphs. The class $(H_1, \ldots, H_k)$-free is the class of all $(H_1, \ldots, H_k)$-free graphs; each class $(H_1, \ldots, H_k)$-free is hereditary.

By $I_t$, $K_t$, $P_t$, and $C_t$ we denote the independent set, the clique, the path, and the cycle on $t$ vertices; $K_{1,t}$ is the claw with $t$ leaves. By $H \mathbin{\dot\cup} H'$ we denote the disjoint union of $H$ and $H'$; we use $tK_2$ for the disjoint union of $t$ graphs $K_2$. By $\overline{G}$ we denote the (edge) complement of $G$. The graph $\overline{K_2 \mathbin{\dot\cup} I_2}$, i.e., the same as a $K_4$ minus one edge, is called diamond.

We recall that GI-completeness is inherited by superclasses while polynomial-solvability of Graph Isomorphism is inherited by subclasses. Also recall that Graph Isomorphism on a class $\mathcal{C}$ is exactly as hard as on $\overline{\mathcal{C}}$, the class of complements of graphs in $\mathcal{C}$. Note that any $H$-free graph is also $H'$-free if $H \leq H'$.

**Proposition 1.** *Let $H_1, H_2$ be graphs and let $\mathcal{C}$ be any hereditary graph class.*

1. $\overline{(H_1, H_2)\text{-free}} = (\overline{H_1}, \overline{H_2})\text{-free}$.

2. $(H_1, H_2)$-free $\subseteq (H'_1, H'_2)$-free *for any $H'_1, H'_2$ with $H_1 \leq H'_1$ and $H_2 \leq H'_2$.*

3. $H_1, H_2 \notin \mathcal{C}$ *implies $\mathcal{C} \subseteq (H_1, H_2)$-free.*

**Definition 1.** *The pan is the graph $\overline{P_3 \mathbin{\dot\cup} K_1}$, i.e., a vertex and triangle joined by one edge. A graph is basic, if it is an independent set, a complete graph, the graph $P_3 \mathbin{\dot\cup} K_1$, or the pan.*

We now show that in the context of the isomorphism problem excluding a basic graph is equivalent to excluding a complete graph or an independent set.

**Lemma 1.** *Let $G$ be a graph that contains $P_4$ as an induced subgraph.*

1. *If $G$ is co-connected then it contains $I_3$ if and only if it contains $P_3 \mathbin{\dot\cup} K_1$.*

2. *If $G$ is connected then it contains $K_3$ if and only if it contains a pan.*

*Proof.* By complementarity it suffices to prove Part 2. Fix a $P_4$ in the graph. Containment of a pan trivially implies containment of a triangle. For the converse, it can be easily verified that there is a pan, if some some triangle contains at least two vertices of the $P_4$. Else, if a triangle contains one vertex $p$ of the $P_4$, we can add a vertex of the $P_4$ adjacent to $p$ to the triangle, obtaining a pan. Else (i.e., if no triangle is incident with the $P_4$) consider the triangle closest to the $P_4$. Due to connectivity, there is a vertex that is adjacent to some vertex of the triangle and closer to the $P_4$. If this vertex is adjacent to exactly one vertex of the triangle, a pan arises. Otherwise we find a closer triangle, which contradicts our initial choice. □

**Theorem 1.** *Graph Isomorphism on a class $\mathcal{C}$ of $K_3$-free graphs is polynomial time equivalent to GI on the subclass of $\mathcal{C}$ that contains all pan-free graphs of $\mathcal{C}$.*

*Proof.* Since Graph Isomorphism for $P_4$-free graphs (so-called cographs) is solvable in polynomial time [7], the theorem follows from Lemma 1 and the fact that graph isomorphism can be solved by comparing connected components. □

## 3 Hardness results

Our standard method to show GI-completeness for Graph Isomorphism on some graph class $\mathcal{H}$ works by reducing the isomorphism problem of a class $\mathcal{H}'$ for which Graph Isomorphism is known to be GI-complete to a subclass of $\mathcal{H}$. For this we require a mapping $\pi \colon \mathcal{H}' \to \pi(\mathcal{H}') \subseteq \mathcal{H}$ which is computable in polynomial time and for which the images of two graphs are isomorphic if and only if the two original graphs are. We call such a mapping $\pi$ a *GI-reduction*. To show hardness for a class $(H_1, H_2)$-free it suffices to provide a GI-reduction $\pi$ for which no graph $G \in \pi(\mathcal{H}')$ contains $H_1$ or $H_2$ as an induced subgraph, implying that $\pi(\mathcal{H}') \subseteq (H_1, H_2)$-free.

Our first reductions prove hardness results for bipartite graphs, split graphs, and line graphs. However, the (previously known) GI-completeness for these particular graph classes is not sufficient. We require hardness for more specific subclasses avoiding specific small graphs. Subsequently, using a more involved reduction, we show that isomorphism of $(P_4 \dot\cup K_1, K_4)$-free graphs is GI-complete.

### 3.1 Bipartite graphs

A straightforward GI-reduction consists of subdividing each edge of a graph by a new vertex. Since the obtained graphs are bipartite, this proves that Graph Isomorphism remains GI-complete on bipartite graphs. This also implies that Graph Isomorphism remains GI-complete on $(H_1, H_2)$-free graphs unless one of the graphs is bipartite, since the class $(H_1, H_2)$-free contains all bipartite graphs if neither $H_1$ nor $H_2$ is bipartite. Let us observe however, that we can draw stronger conclusions namely that Graph Isomorphism remains GI-complete on connected bipartite graphs without induced cycles of length 4, for which the vertices in one of the partition classes have degree two. The following definition allows us to make a first structural observation for the graphs $H_1, H_2$:

**Definition 2.** *A path-star is a subdivision of the $t$-claw $K_{1,t}$, for some $t \in \mathbb{N}$.*

**Lemma 2.** *If neither $H_1$ nor $H_2$ is a disjoint union of path-stars, then Graph Isomorphism on the class $(H_1, H_2)$-free is GI-complete.*

*Proof.* If a graph is not a disjoint union of path-stars, then it either contains two vertices of degree at least 3 which are in the same connected component, or it contains a cycle. We use that two graphs $G_1$ and $G_2$ are isomorphic, if and only if the graphs obtained by subdividing each edge in $G_1$ and $G_2$ respectively are isomorphic. For any integer $c$ there is an integer $c'$ such that a graph that has been subdivided $c'$ times neither contains a cycle of length at most $c$ nor two vertices of degree at least three which are at a distance of at most $c$ apart.

Thus, with a finite number of subdivision steps, we can reduce Graph Isomorphism on general graphs to isomorphism on $(H_1, H_2)$-free graphs. □

Using Part 1 of Proposition 1, we conclude that unless Graph Isomorphism is GI-complete on $(H_1, H_2)$-free, one of the graphs $H_1$, $H_2$ is a forest and one of the graphs is a co-forest.

**Lemma 3.** *A graph $H$ and its complement $\overline{H}$ are forests, if and only if $H \leq P_4$.*

*Proof.* For the if part it suffices to observe that $P_4$ is a self-complementary forest. The only if part is true for graphs of size at most 4. A forest on $n \geq 5$ vertices has at most $n-1$ edges. Since $2(n-1) < \binom{n}{2}$ for $n \geq 5$ the statement follows. □

Graph Isomorphism for $P_4$-free graphs (cographs) is in P [7]. With the previous two lemmas one can conclude that it remains GI-complete on $H$-free graphs when $H$ is not an induced subgraph of the $P_4$; this gives a simple dichotomy for the case of a single forbidden induced subgraph.

**Theorem 2** (see [4]). *Let $H$ be a graph. Graph Isomorphism on $H$-free graphs is in P, if $H \leq P_4$. GI on $H$-free graphs is GI-complete, if $H \not\leq P_4$.*

In the following, we focus on graph classes characterized by two forbidden induced subgraphs. Since isomorphism of $P_4$-free graphs is in P, we assume from now on that $H_1 \not\leq P_4$ and $H_2 \not\leq P_4$. Due to Lemmas 2 and 3 and Part 1 of Proposition 1 we may further assume that $H_1$ is a disjoint union of path-stars and $H_2$ is the complement of a disjoint union of path-stars.

Being forests, unions of path-stars are bipartite. Since bipartite graphs play a repeated role, we introduce some terminology: For a bipartite graph $G$, which has been partitioned into two classes, the *bipartite complement* is the graph obtained by replacing all edges that run between vertices from different partition classes by non-edges and vice versa. (Note that the bipartite complement for unpartitioned bipartite graphs is only well defined if the graph is connected.) A *crossing co-cycle* is a set of vertices that form a cycle in the bipartite complement.

**Lemma 4.** *Isomorphism of $(H_1, H_2)$-free graphs is GI-complete unless $H_1$ or $H_2$ can be partitioned as a bipartite graph without crossing co-cycle.*

*Proof.* Graph Isomorphism is GI-complete on connected graphs. By repeatedly subdividing a connected graph we produce a bipartite graph with an arbitrarily high girth. If at least three subdivisions have been performed on a non-trivial graph, its bipartite complement is connected. Thus taking the bipartite complement of such graphs is a GI-reduction (the bipartite complement of the bipartite complement is the original graph), and we obtain the lemma. □

**Lemma 5.** *For each $G \in \{3K_2, 2K_2 \mathbin{\dot\cup} I_2, P_4 \mathbin{\dot\cup} I_2\}$, Graph Isomorphism on the class on the class of $(H_1, H_2)$-free graphs is GI-complete unless one of the graphs $H_1$ and $H_2$ is bipartite and does not contain the graph $G$.*

*Proof.* Using Lemma 4 this follows, since none of the graphs $3K_2$, $2K_2 \mathbin{\dot\cup} I_2$, and $P_4 \mathbin{\dot\cup} I_2$ can be partitioned as bipartite graph without crossing co-cycle. □

## 3.2 Split graphs

We now turn our attention to split graphs. A *split graph* is a graph whose vertices can be partitioned into an independent set and a clique. Recall that the split graphs are exactly the $(2K_2, C_4, C_5)$-free graphs. The reduction that subdivides each edge and connects all newly introduced vertices produces a split graph, and thus proves GI-completeness on that class. As in the previous section we are able to draw further conclusions about the obtained graphs.

**Definition 3.** Let $G$ be a split graph that has been partitioned into a clique $K$ and an independent set $I$. We say that the partition is

1. of type 1, if the vertices in class $K$ have at most 2 neighbors in class $I$.

2. of type 2, if the vertices in class $K$ have at most 2 non-neighbors in class $I$.

3. of type 3, if the vertices in class $I$ have at most 2 neighbors in class $K$.

4. of type 4, if the vertices in class $I$ have at most 2 non-neighbors in class $K$.

An unpartitioned graph $G$ is said to fulfill split graph condition $i$ (with $i \in \{1, 2, 3, 4\}$), if there is a split partition of the graph that is of type $i$.

**Lemma 6.** *For any $i \in \{1, 2, 3, 4\}$, Graph Isomorphism on the class of graphs that fulfill split graph condition $i$ is GI-complete.*

The proof is deferred to the appendix. Since the class of graphs which fulfill condition $i$ is closed under taking induced subgraphs, the lemma implies that Graph Isomorphism on the class $(H_1, H_2)$-free is GI-complete unless for all $i \in \{1, 2, 3, 4\}$ one of the graphs $H_1$ or $H_2$ fulfills split graph condition $i$.

## 3.3 Line graphs

The next graph class we consider is the class of line graphs. The line graph of a graph $G = (V, E)$ is the graph $L(G) = (E, E')$, in which two vertices are adjacent, if they represent two incident edges in the graph $G$. The line graph of a graph $G$ encodes the isomorphism type of a graph $G$.

**Lemma 7** ([23])**.** *Let $G_1, G_2$ be connected graphs such that neither is a triangle. Then $G_1$ and $G_2$ are isomorphic if and only if their line graphs are.*

The class of line graphs has a characterization by 9 forbidden subgraphs [3]. However, we reduce to a subclass characterized by three forbidden subgraphs.

**Lemma 8.** *Line graphs of graphs of girth at least 5 contain no $K_{1,3}$, no $C_4$, and no diamond.*

*Proof.* A claw $K_{1,3}$ in a line graph $L(G)$ would correspond to three edges in $G$ that each share an endpoint with a fourth edge; then two of the three edges must share an endpoint (forcing an additional edge in $L(G)$). A $C_4$ in $L(G)$ corresponds to a $C_4$ in $G$. Finally, if three edges of a triangle-free graph $G$ pairwise share an endpoint, then they all three share the same endpoint. A fourth edge can therefore not share an endpoint with exactly two of the edges without forming triangle in $G$, i.e., there can be no diamond in $L(G)$. □

Since there is essentially a one-to-one correspondence between graphs and their line graphs, and Graph Isomorphism is GI-complete on triangle-free graphs (since $K_3 \not\leq P_4$), it is also GI-complete on line graphs of triangle-free graphs.

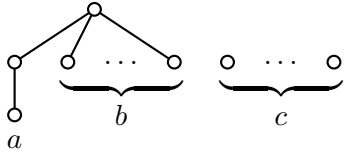

Figure 2: A claw with a subdivision and added isolated vertices. Every split path-star is a subgraph of such a graph. They are denoted by $H(a,b,c)$ where $a$ is the number of subdivided edges, $a+b$ the degree of the claw, and $c$ the number of isolated vertices.

**Lemma 9.** *Graph Isomorphism is Graph Isomorphism-complete on the class of (diamond, claw, $C_4$)-free graphs.*

*Proof.* Since Graph Isomorphism is GI-complete on connected graphs of girth at least 5, and since graphs of girth at least 5 do not contain triangles, the lemma follows from Lemmas 7 and 8. □

### 3.4 A reduction to $(P_4 \mathbin{\dot\cup} K_1, K_4)$-free graphs

In Section B (as a part of the appendix) we discuss a reduction that reduces the class of all graphs into the $(P_4 \mathbin{\dot\cup} K_1, K_4)$-free graphs. Our reduction generalizes the reduction to bipartite graphs, making a replacement of the edges while additionally connecting some of the so-created independent sets. In this sense the reduction (as many other established reductions) is part of a larger scheme of GI-reductions, which use finitely many independent sets, cliques, and relationships between them to encode graphs. We obtain the following theorem.

**Theorem 3.** *Graph Isomorphism is GI-complete on $(P_4 \mathbin{\dot\cup} K_1, K_4)$-free graphs.*

## 4 Structural results and polynomially-solvable cases

We have previously seen that Graph Isomorphism is GI-complete on the $(H_1, H_2)$-free graphs unless each of the four split conditions is fulfilled by one of the two forbidden graphs (Lemma 6). This gives rise to two fundamental cases, namely whether or not one of the two graphs simultaneously fulfills all split conditions. In this section we address the case that neither graph fulfills all split conditions simultaneously. Amongst other conclusions this implies that both graphs must be split graphs or Graph Isomorphism will remain GI-complete. Recall also that one graph must be a path-star while the other must be the complement of a path-star or isomorphism of $(H_1, H_2)$-free graphs will be GI-complete (w.l.o.g. neither of the graphs is an induced subgraph of $P_4$ thus only $H_i$ or $\overline{H}_i$ can be a forest or path-star). Using the results of the previous section we are able to fully characterize this case into classes with either polynomial or GI-complete isomorphism problems.

Without loss of generality we take $H_1$ to be a union of path-stars and $H_2$ to be the complement of a union of path-stars. We analyze first $H_1$; the graph $H_2$ must be a complement of the possible graphs we obtain. Since $H_1$ is split, i.e., $(2K_2, C_4, C_5)$-free, it is a $2K_2$-free path-star. Therefore it contains at most one non-trivial component and no induced path $P_5$ on five vertices. Thus if $H_1$ has a vertex $v$ of degree three or larger, then at most one induced path of length two is emanating from $v$. Together these observations show that $H_1$ is an induced subgraph of the type of graph depicted in Figure 2.

We denote by $H(a,b,c)$ the graph that is depicted in Figure 2, with $a \in \{0,1\}$, $b \in \mathbb{N}$, and $c \in \mathbb{N}$. We require that if $a = 1$ then $b > 0$, otherwise (if $a = 1$ and $b = 0$) we can reinterpret the graph with $b = 2$ and $a = 0$ (thus, $a = 1$ iff the graph contains a $P_4$). We also require $a+b \geq 1$ since the independent set and the clique fulfill all split conditions. Observe that any induced subgraph of some $H(a, b, c)$ is isomorphic to $H(a', b', c')$ for some values of $a', b', c'$ (it suffices to consider the induced subgraphs of the claw with one subdivision). We will argue that under these restrictions we may focus on the case that $a = a' = 0$, since Graph Isomorphism remains GI-complete otherwise.

**Lemma 10.** *Let $H_1 = H(a, b, c)$ and $H_2 = \overline{H(a', b', c')}$ such that neither graph fulfills all split conditions and such that $a + a' \geq 1$. Then GI remains GI-complete on $(H_1, H_2)$-free graphs.*

*Proof.* W.l.o.g. $a = 1$ and $b \geq 1$ (else complement and switch $H_1$ and $H_2$). Since $P_4 \mathbin{\dot{\cup}} K_1$, its induced subgraphs, and the respective complements fulfill all split conditions, by assumption, $H_1, \overline{H}_2 \not\leq P_4 \mathbin{\dot{\cup}} K_1$, i.e., $H_1, \overline{H}_2 \notin \{P_4 \mathbin{\dot{\cup}} K_1, P_4, P_3 \mathbin{\dot{\cup}} K_1, P_3, K_2 \mathbin{\dot{\cup}} I_2, K_2 \mathbin{\dot{\cup}} K_1, K_2\}$, corresponding to forbidden triples

$$(a, b, c), (a', b', c') \notin \{(1, 1, 1), (1, 1, 0), (0, 2, 1), (0, 2, 0), (0, 1, 2), (0, 1, 1), (0, 1, 0)\}.$$

Now if $c \geq 2$ then $P_4 \mathbin{\dot{\cup}} I_2 \leq H_1$, implying that $H_1$ is not a bipartite graph without crossing co-cycle. Thus Graph Isomorphism remains GI-complete by Lemma 4, since $H_2$ is not bipartite (by $a' + b' \geq 1$ and the forbidden triples we have $a' + b' + c' \geq 3$, i.e., $I_3 \leq \overline{H}_2$). For the remaining part we assume that $c \leq 1$. Since $(a, b, c) \notin \{(1, 1, 0), (1, 1, 1)\}$ we may conclude $b \geq 2$, and therefore we have that $K_{1,3} \leq H(1, 2, 0) \leq H_1$ and that $K_2 \mathbin{\dot{\cup}} I_2 \leq H(1, 2, 0) \leq H_1$.

If $a' + b' \geq 3$ then Graph Isomorphism remains GI-complete on $(H_1, H_2)$-free, by Lemma 9, since $H_2 = \overline{H(a', b', c')}$ with $a' + b' \geq 3$ contains the co-claw. The remaining case is that $a' + b' \leq 2$, i.e., $(a', b') \in \{(1, 1), (0, 2), (0, 1)\}$. Considering all excluded triples, e.g., $(1, 1, 1)$ and $(1, 1, 0)$, we conclude that $c' \geq 2$. Thus $K_2 \mathbin{\dot{\cup}} I_2 \leq \overline{H}_2$, i.e., $diamond \leq H_2$. Thus Graph Isomorphism is GI-complete on $(K_{1,3}, diamond)$-free $\subseteq (H_1, H_2)$-free graphs by Lemma 9. □

For the remaining discussion we may thus assume that $a = 0$ and $a' = 0$.

**Theorem 4.** *Isomorphism of $(H(0, b, c), \overline{H(0, b', c')})$-free graphs is in P when:*

1. *$b = 0$ or $b' = 0$ (i.e., one of the graphs is a clique or an independent set),*

2. *$c, c' \leq 1$ and $b, b' \geq 1$,*

3. *$c, c' \geq 2$ and $b, b' \in \{1, 2\}$,*

4. *$(c \geq 2, c' \leq 1, b \geq 1, b' \in \{1, 2\})$, or $(c' \geq 2, c \leq 1, b' \geq 1, b \in \{1, 2\})$.*

*In all other cases it is GI-complete.*

To prove the theorem we use vertex-colorings of the input graphs. (In the context of Graph Isomorphism the vertex colorings are not assumed to be proper). We say that a vertex-colored graph has *bounded color valences*, if there is a constant $D$, such that for every color class $C$ every vertex $v$ (possibly in $C$) has at most $D$ neighbors or at most $D$ non-neighbors in $C$. We will show (Lemma 14) that in a graph without $H(0, b, c)$ and $\overline{H(0, b', c')}$ bounded color valence within color classes implies bounded color valence overall. Bounding the color valence one can reduce the isomorphism problem to that of graphs of bounded degree.

**Theorem 5** (Babai, Luks [1]). *Graph Isomorphism for colored graphs of bounded color valence is solvable in polynomial time.*

To prove Theorem 4 we distinguish cases according to the numbers $c$ and $c'$ in the forbidden subgraphs $H(0, b, c)$ and $\overline{H(0, b', c')}$.

*General proof strategy for Theorem 4.* We defer the proof of Theorem 4 to Section C of the appendix and instead provide a high level description of the general proof-strategy. When proving each of the four cases our strategy is as follows: The starting observation (Lemma 14) is that a colored $(H(0, b, c), \overline{H(0, b', c')})$-free graph, which has bounded degree or bounded co-degree within each color class, also has bounded color valence between different color classes. This enables the use of Theorem 5. Thus, we now intend to find a canonical (in particular isomorphism-respecting) way of coloring both input graphs, so that the color classes have bounded degree or bounded co-degree. We employ two methods to pick color classes, both of them ensure that the coloring preserves isomorphism. Either we choose the colors of the vertices by properties of the vertices that can be computed in polynomial time. Or we guess an ordered set of vertices of constant size, color the vertices in this set with singleton colors, and then color the remaining vertices according to their adjacencies to the vertices in the ordered set. Guessing a constant number $k$ of vertices increases the running time by a factor of $n^k$, and can therefore be performed in polynomial time. The second coloring operation is typically referred to as individualization.

In Case 1 we individualize one vertex, and use induction to obtain a canonical coloring with the desired properties. In Case 2 we individualize one vertex, and use a combinatorial argument to show that this gives a canonical coloring with the desired properties. In Case 4, using Lemma 1, we reduce the problem to $(H(0, b, c), K_3)$-free graphs, and then apply induction on $c$ to obtain the canonical coloring. Case 3 is the most interesting one (and rather involved). In this case, by individualizing a finite number of vertices, we can obtain a colored graph, in which each of the color classes is a cluster, or a co-cluster graph. (A cluster is a $P_3$-free graph or equivalently a disjoint union of cliques.) For our purpose this is not sufficient, as for example a cluster graph can have vertices that simultaneously have large degree and large co-degree. We call a cluster $d$-diverse if it contains at least $d$ disjoint cliques of size at least $d$. A $d$-diverse co-cluster is the complement of a $d$-diverse cluster. We show (Lemma 18) that for large $d$ a $(H(0, 2, c), \overline{H(0, 2, c')})$-free graph cannot contain a $d$-diverse cluster and a $d$-diverse co-cluster at the same time. With this (possibly taking complements) our situation simplifies to the case where there is one color class $A$ that is a cluster, and all other color classes are of bounded degree or bounded co-degree. After splitting off a bounded number of cliques from $A$, we can show that for each of the remaining cliques there is only a bounded number of types by which the vertices are connected to the vertices outside the cluster. Using this we replace the cluster by a bounded number of representatives, one for each type, color-encoding the number of vertices of each type. This leaves a graph with bounded color valence and enables us to apply Theorem 5. □

## 5 The remaining cases

In the previous section we investigated the case when neither of the two forbidden induced subgraphs fulfills all split graph conditions. We now consider the case, where one of the two graphs fulfills all split graph conditions. W.l.o.g. we let $H_1$ be this graph and require that $H_1$ is a disjoint union of path-stars (otherwise we take complements); there are only few choices for $H_1$. The proofs of Lemma 11 and Theorem 6 can be found in Appendix D.

**Lemma 11.** *If a graph $G$ is a union of path-stars and fulfills all split graph conditions, then it is an induced subgraph of one of the following graphs (depicted in Figure 1): An independent set, $P_4 \dot\cup K_1$, $K_{1,3} \dot\cup I_2$, or the fork.*

**Theorem 6.** *Suppose $H_1$ is a nonbasic disjoint union of path-stars and fulfills all split graph conditions. If $H_2$ has more than 7 vertices, then an application of one of Lemmas 2, 3, 5, or 9, or*

*Theorems 2, 3, or 4 determines that $(H_1 \mathbin{\dot{\cup}} H_2)$-free is GI-complete or polynomial-time solvable. More strongly, this can be concluded unless $H_1$ is one of the graphs $\{P_4 \mathbin{\dot{\cup}} K_1, K_2 \mathbin{\dot{\cup}} I_2, P_3 \mathbin{\dot{\cup}} I_2\}$ and $\overline{H_2}$ has at most 7 vertices and is a disjoint union of at most 3 paths.*

## 6 Conclusion

In order to initiate a systematic study of the Graph Isomorphism Problem on hereditary graph classes we considered graph classes characterized by two forbidden induced subgraphs. We presented an almost complete characterization of the case that neither of the two forbidden subgraphs is basic into GI-complete and polynomial cases, leaving only few pairs of forbidden subgraphs. Theorem 4 constitutes the main technical contribution towards this result. Together with the tractability of $P_4$-free graphs (Theorem 2, [7]) it establishes the polynomially solvable cases. On the other hand suppose $H_1$ and $H_2$ are non-basic and $(H_1, H_2)$-free is not a polynomial-time solvable case of Theorems 2 or 4. Then, Graph Isomorphism on the class of $(H_1, H_2)$-free graphs is GI-complete, unless for $H_1$ and $H_2$, or for $\overline{H_1}$ and $\overline{H_2}$, one of the graphs is in $\{P_4 \mathbin{\dot{\cup}} K_1, K_2 \mathbin{\dot{\cup}} I_2, P_3 \mathbin{\dot{\cup}} I_2\}$, and the other graph has at most 7 vertices and is the complement of a union of at most 3 paths.

Several further cases, e.g., all cases involving the $\overline{P_6}$ or the $\overline{P_7}$, can be excluded by variants of the reduction used for Theorem 3. Of the remaining cases, in a preprint, Rao [19] resolves positively the case $(P_4 \mathbin{\dot{\cup}} K_1, \overline{P_4 \mathbin{\dot{\cup}} K_1})$-free and its subclasses; similar (modular) decomposition techniques appear to apply to other cases as well. Several of the remaining cases are classes of bounded clique-width [6], which could indicate their tractability.

For the case in which one of the forbidden graphs is basic, our reductions and our polynomial-time algorithms are still applicable and resolve a large portion of the cases. However, as mentioned in the introduction, complete resolution appears to require new techniques. Future steps for studying the hereditary graph classes include the resolution of the remaining cases and analysis of graph classes characterized by more than two forbidden subgraphs.

## References

[1] L. Babai and E. M. Luks. Canonical labeling of graphs. In *STOC '83*, pages 171–183, 1983.

[2] L. Babel, I. N. Ponomarenko, and G. Tinhofer. The isomorphism problem for directed path graphs and for rooted directed path graphs. *Journal of Algorithms*, 21(3):542–564, 1996.

[3] L. W. Beineke. Characterizations of derived graphs. *Journal of Combinatorial Theory, Series B*, 9(2):129–135, 1970.

[4] K. S. Booth and C. J. Colbourn. Problems polynomially equivalent to graph isomorphism. Technical Report CS-77-04, Comp. Sci. Dep., Univ. Waterloo, 1979.

[5] R. B. Boppana, J. Hastad, and S. Zachos. Does co-NP have short interactive proofs? *Information Processing Letters*, 25:127–132, 1987.

[6] A. Brandstädt, F. F. Dragan, H. Le, and R. Mosca. New graph classes of bounded clique-width. *Theory of Computing Systems*, 38:623–645, 2005.

[7] D. G. Corneil, H. Lerchs, and L. Stewart Burlingham. Complement reducible graphs. *Discrete Applied Mathematics*, 3(3):163–174, 1981.

[8] M. Grohe and D. Marx. Structure theorem and isomorphism test for graphs with excluded topological subgraphs. In H. J. Karloff and T. Pitassi, editors, *STOC*, pages 173–192. ACM, 2012.

[9] W. L. Hsu. $O(m \cdot n)$ algorithms for the recognition and isomorphism problems on circular-arc graphs. *SIAM Journal on Computing*, 24(3):411–439, 1995.

[10] J. Köbler, U. Schöning, and J. Torán. *The graph isomorphism problem: its structural complexity*. Birkhäuser Verlag, Basel, Switzerland, Switzerland, 1993.

[11] D. Král, J. Kratochvíl, Z. Tuza, and G. J. Woeginger. Complexity of coloring graphs without forbidden induced subgraphs. In *WG '01*, pages 254–262, 2001.

[12] S. Kratsch and P. Schweitzer. Isomorphism for graphs of bounded feedback vertex set number. In *SWAT*, volume 6139 of *LNCS*, pages 81–92, 2010.

[13] V. V. Lozin. A decidability result for the dominating set problem. *Theoretical Computer Science*, 411(44–46):4023–4027, 2010.

[14] G. S. Lueker and K. S. Booth. A linear time algorithm for deciding interval graph isomorphism. *Journal of the ACM*, 26(2):183–195, 1979.

[15] E. M. Luks. Isomorphism of graphs of bounded valence can be tested in polynomial time. *Journal of Computer and System Sciences*, 25(1):42–65, 1982.

[16] S. Nakano, R. Uehara, and T. Uno. A new approach to graph recognition and applications to distance-hereditary graphs. *Journal of Computer Science and Technology*, 24:517–533, 2009.

[17] I. N. Ponomarenko. The isomorphism problem for classes of graphs closed under contraction. *Journal of Mathematical Sciences*, 55(2):1621–1643, 1991.

[18] F. P. Ramsey. On a problem of formal logic. *Proceedings of the London Mathematical Society*, s2-30(1):264–286, 1930.

[19] M. Rao. Decomposition of (gem,co-gem)-free graphs. unpublished, available at http://www.labri.fr/perso/rao/publi/decompgemcogem.ps.

[20] U. Schöning. Graph isomorphism is in the low hierarchy. *Journal of Computer and System Sciences*, 37(3):312–323, 1988.

[21] P. Schweitzer. *Problems of unknown complexity: Graph isomorphism and Ramsey theoretic numbers*. PhD thesis, Universität des Saarlandes, Germany, 2009.

[22] R. Uehara, S. Toda, and T. Nagoya. Graph isomorphism completeness for chordal bipartite graphs and strongly chordal graphs. *Discrete Applied Mathematics*, 145(3):479–482, 2005.

[23] H. Whitney. Congruent graphs and the connectivity of graphs. *American Journal of Mathematics*, 54(1):150–168, 1932.

# A  Proof of Lemma 6

*Proof.* We show the statement for condition 1. For a graph $G$ we define $SP1(G)$ to be the graph obtained by subdividing each edge, and then connecting all newly added vertices. Two graphs $G_1$ and $G_2$ with $n$ vertices and $m$ edges and $|m - n| > 1$ are isomorphic, if and only if $SP1(G_1)$ and $SP1(G_2)$ are isomorphic: The one direction follows from the construction. For the other direction suppose $SP1(G_1)$ and $SP1(G_2)$ are isomorphic. We partition the vertices of $SP1(G_1)$ and $SP1(G_2)$ into a split decomposition, such that the independent set contains $n$ vertices and the clique contains $m$ vertices. Since $|m - n| > 1$ this decomposition is equal to the decomposition into original vertices and vertices added by the subdivision: Indeed, since the original vertices form an independent set, at most one of these vertices can be contained in the clique and since the added vertices form a clique at most one of them can be added to the independent set. Thus every added vertex has at least one neighbor among the original vertices in the independent set; hence, there is only one possible partition, with partition classes of size $n$ and $m$.

Given two graphs $G_1$ and $G_2$ we may solve Graph Isomorphism in the following way: If the number of vertices or the number of edges differs we report non-isomorphism. Otherwise, suppose $n$ is the number of vertices and $m$ is the number of edges of either graph. W.l.o.g. $n > 3$. If the number of edges differs from the number of vertices by at most one, we add a universal vertex. This changed $n$ by 1 and $m$ by more than 3. We then transform the graphs with the $SP1(\cdot)$-operation and solve Graph Isomorphism.

The proof for the other conditions follows from complementation arguments: For the reduction for condition 2 we complement all edges running between the clique and the independent set. The other conditions are obtained by complementing the entire graph. □

# B  A reduction to $(P_4 \mathbin{\dot\cup} K_1, K_4)$-free graphs

In this section we discuss a GI-reduction that leads to a proof of Theorem 3 from Section 3.4. After introducing the reduction we first prove that it is indeed a GI-reduction and then proceed to show that its image is $(P_4 \mathbin{\dot\cup} K_1, K_4)$-free; together this constitutes a proof of the theorem.

**Definition 4.** We define $\pi := \pi(1\cdot\cdot 2\cdot\cdot 3\cdot\cdot 4\cdot\cdot 5; 135, 24) \colon \mathcal{G} \to \mathcal{G}$ to be the mapping that alters any graph $G$ by performing the following steps (see Figure 3):

1. Add five universal vertices to $G$ and connect them. Abusing notation we call the obtained graph $G$, which is now surely connected and has minimum degree four.

2. Subdivide each edge of $G$ and let $Q$ denote the set of added vertices (of degree two).

3. Add a vertex that is adjacent to all vertices of $Q$. Now vertices in $Q$ have degree three. Let $P$ be the set of original vertices plus the newly added vertex. Note that vertices in $P$ have degree at least four and that the current graph is bipartite.

4. Replace each edge $e = \{p, q\}$, with $p \in P$ and $q \in Q$, by edges $\{p, a_e\}$, $\{a_e, b_e\}$, $\{b_e, c_e\}$, and $\{c_e, q\}$, where $a_e$, $b_e$, and $c_e$ are new vertices. Let $A$, $B$, and $C$ be the sets of new vertices $a_e$, $b_e$, respectively $c_e$.

5. Complement all edges between $P$ and $A$, $A$ and $B$, $B$ and $C$, as well as between $C$ and $Q$.

6. Add all edges between $P$ and $B$, $P$ and $Q$, $B$ and $Q$, as well as between $A$ and $C$.

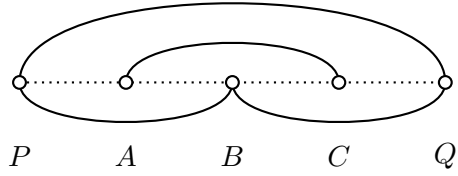

Figure 3: A schematic representation of the reduction to $(P_4 \dot\cup K_1, K_4)$-free graphs.

Observe that each edge between $P$ and $Q$ is now encoded as a path of non-edges from $P$ over $A$, $B$, and $C$ to $Q$. This as well as the complete adjacency between some of the sets, see Figure 3, are the main ingredients. Let us also point out some further properties:

1. The vertex sets $P$, $A$, $B$, $C$, and $Q$ are independent sets.

2. Each vertex in $A$ has exactly one non-neighbor in $P$ and one non-neighbor in $B$. The same is true for vertices in $B$ towards $A$ and towards $C$ as well as for vertices in $C$ towards $B$ and towards $Q$. Vertices in $Q$ have exactly three non-neighbors in $C$ and vertices of $P$ have at least four non-neighbors in $A$.

Now we can prove that $\pi$ is indeed a GI-reduction and, subsequently, that it generates only $(P_4 \dot\cup K_1, K_4)$-free graphs. Together these facts prove Theorem 3, namely that Graph Isomorphism remains GI-complete on $(P_4 \dot\cup K_1, K_4)$-free graphs.

**Lemma 12.** *The mapping $\pi$ is a GI-reduction.*

*Proof.* Let $G_1$ and $G_2$ be two graphs, w.l.o.g. they have the same numbers of vertices and edges respectively.

It is easy to see that any isomorphism $\phi : V(G_1) \to V(G_2)$ can be extended to an isomorphism of $\pi(G_1)$ and $\pi(G_2)$: First extend it to the universal vertices. Then, when an edge $\{u, v\}$ is subdivided by a vertex $q$, let $\phi(q)$ be the vertex that subdivides $\{\phi(u), \phi(v)\}$. Observe that this gives an isomorphism of the graphs obtained after Step 2. Extend $\phi$ to include the vertex added in Step 3. For the triple subdivision of edges $e = \{p, q\}$ in Step 4, let the images of $a_e$, $b_e$, and $c_e$ be the corresponding vertices that subdivide $\{\phi(p), \phi(q)\}$. At this point $\phi$ is an isomorphism of the graphs obtained after Step 4 that maps the vertex sets $P$, $A$, $B$, $C$, and $Q$ of $G_1$ to their corresponding counterparts in $G_2$. Thus complementation of adjacency between some of the sets respectively adding all edges between pairs of sets does not influence $\phi$, implying that $G'_1$ and $G'_2$ are isomorphic.

Now for the converse let $\phi$ be an isomorphism of $\pi(G_1)$ and $\pi(G_2)$ for two graphs $G_1$ and $G_2$. Let $n$ denote the number of vertices after Step 1 and let $m$ denote the number of edges; note that minimum degree four implies $m \geq 2n$. Thus Step 2 creates a set $Q$ of size $m$ and doubles the edges to $2m$. The set $P$, created in Step 3, contains the original $n$ vertices with a degree between four and $n - 1$ as well as the universal vertex of degree $|Q| = m$; i.e., $|P| = n + 1$. The vertices in $Q$ have degree exactly three after Step 3, and hence there are $3m$ edges. Therefore the sets $A$, $B$, and $C$, created in Step 4 each have size $3m$. Let us consider the co-degrees of the vertices in $\pi(G_1)$ and $\pi(G_2)$:

**P** The $n$ original vertices in $P$ have at least four non-neighbors in $A$, they are not adjacent to other vertices of $P$ or vertices of $C$. Thus their co-degree is at least $4 + (|P| - 1) + |C| = n + 3m + 4$. We can upper bound the co-degree since those vertices originally have degree at most $n - 1$, i.e., they have at most $n - 1$ neighbors in $Q$ before Step 4. This fact bounds the number of non-neighbors in $A$ by $n - 1$, giving an upper bound of $(n-1) + (|P|-1) + |C| = 2n + 3m - 1$. Finally, the vertex adjacent to all of $Q$ has degree $m$ before Step 4, making its co-degree $m + (|P|-1) + |C| = n + 4m$.

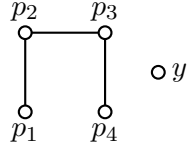

Figure 4: A $P_4 \mathbin{\dot\cup} K_1$ on vertices $p_1,\ldots,p_4$ and $y$ as used in Lemma 13.

**A** These vertices are not adjacent to $Q$, completely adjacent to $C$, and have exactly one non-neighbor each in $P$ and $B$. Since $A$ is an independent set, their co-degree is exactly $|Q| + 2 + (|A|-1) = 4m + 1$.

**B** These vertices are completely adjacent to $P$ and $Q$, have no edges among them, and have exactly one non-neighbor each in $A$ and $C$. Their co-degree is exactly $(|B|-1) + 2 = 3m + 1$.

**C** Vertices of $C$ are not adjacent to $P$, completely adjacent to $A$, and have no edges connecting themselves. Furthermore they have exactly one non-neighbor each in $B$ and $Q$, making their co-degree exactly $|P| + (|C|-1) + 2 = n + 3m + 2$.

**Q** Each vertex of $Q$ has exactly three non-neighbors in $C$. Furthermore it is completely connected to $P$ and $B$ and there are no edges to $A$. Since $Q$ is an independent set, each vertex has co-degree exactly $3 + |A| + (|Q|-1) = 4m + 2$.

Using $m \geq 2n$ we get that the co-degree of a vertex determines to which of the five independent sets it belongs:

$$3m + 1 < n + 3m + 2 < n + 3m + 4 < 2n + 3m - 1 < 4m + 1 < 4m + 2 < n + 4m.$$

This implies that $m$, $n$ and therefore also the correct way to partition the vertices into the five independent sets can be reconstructed (the co-degree values respectively correspond to: $B$, $C$, lower and upper bound for the original vertices in $P$, $A$, $Q$, and the universal vertex in $P$).

Thus any isomorphism $\phi$ of $\pi(G_1)$ and $\pi(G_2)$ must respect the independent sets, and using $\phi$ an isomorphism for $G_1$ and $G_2$ can be found. For this observe that $\phi$ can be restricted to an isomorphism of the bipartite graphs obtained after Step 3, since the edge-replacement of Step 4 can be easily reversed, given the partitions. □

Since $\pi$ is a GI-reduction we conclude that isomorphism is GI-complete on $\pi(\mathcal{G})$. We now analyze $\pi(\mathcal{G})$ and prove that it is $(P_4 \mathbin{\dot\cup} K_1, K_4)$-free. Thus, Lemma 12 and the following Lemma 13 constitute a proof of Theorem 3.

**Lemma 13.** *The image $\pi(\mathcal{G})$ of $\mathcal{G}$ is contained in $(P_4 \mathbin{\dot\cup} K_1, K_4)$-free.*

*Proof.* Consider any graph $G' = \pi(G) = (P \mathbin{\dot\cup} A \mathbin{\dot\cup} B \mathbin{\dot\cup} C \mathbin{\dot\cup} Q, E')$. We first observe that $G'$ is $K_4$-free since it can be partitioned into three independent sets, namely $P \cup C$, $A \cup Q$, and $B$. Let us now assume for contradiction that $G'$ contains $P_4 \mathbin{\dot\cup} K_1$ as an induced subgraph.

We make a case distinction based on which of the vertex sets $P$ through $Q$ contains the isolated vertex $y$. As we will not use the asymmetric degree properties obtained in Step 2 of the reduction, by symmetry it suffices to consider $y \in P$, $y \in A$, and $y \in B$.

**$y \in P$:** Since $\pi$ adds all edges between $P$ and $B$ as well as between $P$ and $Q$ the vertices of the $P_4$ must be contained in $P$, $A$, and $C$. Since $G'[P \cup A \cup C]$ is bipartite (with $A$ as one bipartition class) exactly two vertices of the $P_4$ must be in $A$, namely $p_1$ and $p_3$ or $p_2$ and $p_4$; by symmetry we assume $p_1, p_3 \in A$. Let us now observe that neither $P$ nor $C$ can contain $p_4$: If $p_4 \in P$ then $p_1$ would have two non-neighbors in $P$, namely $p_4$ and $y$. If $p_4 \in C$ then there would be an edge between $p_4$ and $p_1$ since $A$ and $C$ are completely connected.

**$y \in A$:** All vertices of the $P_4$ must be contained in $P$, $A$, $B$, and $Q$ since $y$ is adjacent to all vertices of $C$. By definition of $\pi$ the vertex sets $P$, $B$, and $Q$ form a complete tripartite graph. Hence only two of them can contain vertices of the $P_4$. Furthermore $y$ has exactly one non-neighbor each in $P$ and $B$, thus these sets each contain at most one vertex of the $P_4$. Again none of the sets can contain more than two vertices of the $P_4$.

- If $P$ and $B$ contain vertices of the $P_4$ then each contains exactly one and they must be consecutive in the $P_4$. The set $A$ thus contains two non-adjacent vertices of the $P_4$. Thus $p_1, p_4 \in A$ and w.l.o.g. $p_2 \in P$ and $p_3 \in B$. This however would mean that $p_3 \in B$ has two non-neighbors in $A$, namely $p_1$ and $y$.

- If $P$ and $Q$ contain vertices of the $P_4$, i.e., $B$ contains none, then the $P_4$ is contained in the bipartite graph $G'[P \cup A \cup Q]$ with bipartition $P$ and $A \cup Q$. Thus two vertices must be contained in $P$ but this would imply two non-neighbors in $P$ for $y \in A$.

- If $Q$ and $B$ contain vertices of the $P_4$, i.e., $P$ contains none, then we are considering the bipartite graph $G'[A \cup B \cup Q]$ with bipartition $B$ and $A \cup Q$. Similarly to the previous case this forces $B$ to contain two vertices of the $P_4$ implying that $y \in A$ would have two non-neighbors in $B$.

**$y \in B$:** In this case all vertices of the $P_4$ must be contained in $A$, $B$, and $C$, since $B$ is completely connected to $P$ and $Q$. Both $A$ and $C$ may each contain at most one vertex of the $P_4$ since $y$ may have at most one non-neighbor in either set. Also $B$ can contain at most two vertices of the $P_4$ since it is an independent set. Thus both $A$ and $C$ must each contain one vertex of the $P_4$ and those vertices must be connected, i.e., $p_1$ and $p_2$, $p_2$ and $p_3$, or $p_3$ and $p_4$. Since $B$ can contain only non-adjacent vertices of the $P_4$ we conclude that $p_1, p_4 \in B$ and w.l.o.g. (by symmetry) $p_2 \in A$ and $p_3 \in C$. Then however $p_2$ would contradict the restriction that each vertex in $A$ has at most one non-neighbor in $B$, since it has non-neighbors $p_4$ and $y$ (note that the same is true for $p_3$ and non-neighbors $p_1$ and $y$). □

## C  Proof of Theorem 4

Here we give a complete proof for Theorem 4. We begin the proof by arguing GI-completeness for the classes $(H(0, b, c), \overline{H(0, b', c')})$-free not covered by the four polynomial cases of Theorem 4. We apply Lemma 9, i.e., GI-completeness of Graph Isomorphism on (*diamond*, *claw*)-free: By Case 1 we have $b, b' \geq 1$. Thus if $c \geq 2$ and $b' \geq 3$ then $H(0, b, c)$ would contain the co-diamond and $\overline{H(0, b', c')}$ would contain the co-claw, making Graph Isomorphism GI-complete. The same is true when $c' \geq 2$ and $b \geq 3$ (with diamond and claw). Observe that this establishes hardness for all choices of $b$, $b'$, $c$, and $c'$ not covered by the four polynomial cases of Theorem 4.

The four polynomial cases are proved Sections C.1, C.2, C.3, and C.4 respectively. We will repeatedly require the following lemma on the structure of $H(0, b, c)$-free graphs:

**Lemma 14.** *Let $G$ be a vertex-colored graph that is $H(0, b, c)$-free, and the subgraph induced by the vertices in color red have a degree bounded by $d$. Then for any non-red vertex either the degree towards the red color class (i.e., the number of red neighbors), or the co-degree towards the red color class (i.e., the number of red non-neighbors) is bounded by a function of $d, b$ and $c$. More specifically either the degree or the co-degree is at most $(d + 1) \cdot (b + c)$.*

*Proof.* Let $v$ be a non-red vertex. Suppose for contradiction, that the degree and the co-degree of $v$ is larger than $(d + 1) \cdot (b + c)$. Let $N_{\text{red}}(v)$ be the red neighbors of $v$, and $co\text{-}N_{\text{red}}(v)$ be the red non-neighbors of $v$. It suffices to show that there is an independent set that contains $b$ vertices from $N_{\text{red}}(v)$ and $c$ vertices from $co\text{-}N_{\text{red}}(v)$. By repeatedly picking a vertex from one of the sets, and discarding its neighbors among both sets, we construct the desired independent set. □

We say that a vertex-colored graph has *bounded color valences*, if there is constant $D$, such that for every color class $C$ every vertex $v$ (possibly in $C$) has at most $D$ neighbors or at most $D$ non-neighbors in $C$. The lemma implies that in a graph without $H(0, b, c)$ and $\overline{H(0, b', c')}$ bounded color valence within color classes implies bounded color valence overall. Bounding the color valence one can reduce the isomorphism problem to that of graphs of bounded degree.

**Theorem 7** (Babai, Luks [1]). *Graph Isomorphism for colored graphs of bounded color valence is solvable in polynomial time.*

We will now consider the four cases of Theorem 4.

## C.1 Case I: $b = 0$ or $b' = 0$

In this section we consider the case that $b' = 0$, i.e., $H_1 = H(0, b, c)$ and $H_2 = \overline{H(0, 0, c')} = K_{c'+1}$, and prove that Graph Isomorphism is polynomial-time decidable on $(H_1, H_2)$-free.

**Lemma 15.** *Graph Isomorphism of $(H(0, b, c), \overline{H(0, 0, c')})$-free graphs is in P.*

*Proof.* We show by induction on $c$ and $c'$ that by individualizing a bounded number of vertices, we obtain color classes of bounded degree or bounded co-degree. Using Lemma 14 and applying Theorem 7 we then obtain a polynomial time algorithm. The induction base is $c = 0$ or $c' = 0$. The case $c' = 0$ is trivial: No non-trivial graph is $H(0, 0, 0)$-free.

The case $c = 0$ is the case where $H_1 = K_{1,b}$ and $H_2 = K_{c'+1}$. For such graphs, we have shown a bound on the maximum degree at the beginning of the introduction using a Ramsey argument. For the general case, where $c, c' \geq 1$, it suffices to observe the following: When picking an arbitrary vertex $v$, the rest of the graph decomposes into two subgraphs, the neighbors of $v$ and the non-neighbors of $v$. The neighbors of $v$ form an $\overline{H(0, 0, c' - 1)}$-free graph, and the non-neighbors of $v$ form an $H(0, b, c - 1)$-free graph. The statement now follows by induction. □

## C.2 Case II: $c, c' \leq 1$ and $b, b' \geq 1$

In this section we consider the case $H_1 = H(0, b, 1) = K_{1,b} \dot\cup K_1$ and $\overline{H_2} = \overline{H(0, b', 1)} = K_{1,b'} \dot\cup K_1$ and prove that Graph Isomorphism is polynomial-time decidable on $(H_1, H_2)$-free. We require the following lemma:

**Lemma 16.** *Let $G$ be a connected graph without an $\overline{H(0, b, 1)}$ which contains a clique $K = K_{2 \cdot b + 1}$ of size $(2 \cdot b + 1)$, then every vertex is a neighbor of at least half of the vertices of the clique.*

*Proof.* It suffices to show that every neighbor of a vertex in $K$ is a neighbor of at least $b$ vertices of $K$, and that there is no vertex that has distance 2 from the clique. The first statement follows from the fact that the $\overline{H(0,b,1)}$ is forbidden. The second follows from the first. □

**Lemma 17.** *Graph Isomorphism of $(H(0,b,1), \overline{H(0,b',1)})$-free graphs is in* P.

*Proof.* W.l.o.g. the input graphs and their complements are connected. Let $v$ be an arbitrary vertex of an $(H_1, H_2)$-free graph. Consider $G'$, the neighborhood of $v$ and $G''$ the non-neighborhood of $v$. It suffices to show that $G'$ and $G''$ are both of bounded degree or bounded co-degree. By complementary symmetry it suffices to show this for $G''$. Note that $G''$ does not contain the complete bipartite graph $K_{1,b}$. If $G''$ contains a vertex of degree $R(b, 2 \cdot b' + 1)$ then it contains a clique of size $2 \cdot b' + 1$ (here $R(\cdot, \cdot)$ refers to the Ramsey Number). By Lemma 16 every vertex of $G$, and in particular every vertex of $G''$ is a neighbor of a vertex of the clique. If follows that $G''$ has no independent set of size $2 \cdot b' + 1 \cdot b + 1$, otherwise the vertices in this independent set have a common neighbor in the clique. Finally suppose a vertex $v$ has $c$ non-neighbors, then there is an clique of size $2 \cdot b + 1$ among them. By Lemma 16 one of its vertices is adjacent to $v$, a contradiction. □

## C.3 Case III: $c, c' \geq 2$ and $b, b' \in \{1, 2\}$

In this section we assume that $c, c' \geq 2$ and that $b, b' \in \{1, 2\}$. For these parameters we show that for all $\boldsymbol{c} \in \mathbb{N}$ the isomorphism problem on the class of $(H(0,2,\boldsymbol{c}), \overline{H(0,2,\boldsymbol{c})})$-free graphs is polynomial-time decidable. This establishes tractability on any class $(H(0,b,c), \overline{H(0,b',c')})$-free, where $b, b' \leq 2$ and $c, c' \in \mathbb{N}$, since the class $(H(0,b,c), \overline{H(0,b',c')})$-free is contained in the class $(H(0,2,\boldsymbol{c}), \overline{H(0,2,\boldsymbol{c})})$-free, with $\boldsymbol{c} = \max(c, c')$.

Intuitively, due to the structure of the graphs $H(0,2,\boldsymbol{c})$ and $\overline{H(0,2,\boldsymbol{c})}$ we can individualize a finite number of vertices such that the evolving color classes are either a cluster or a co-cluster graph. For this consider a class of vertices that have the same adjacency to $2\boldsymbol{c}$ selected vertices, then they share $\boldsymbol{c}$ non-neighbors, i.e., the neighborhood must be $P_3$-free (a cluster), or they share $\boldsymbol{c}$ neighbors and must be $\overline{P_3}$-free (a co-cluster).

We distinguish the type of cluster graph that may arise: Let $d$ be an integer. We say that a graph contains a *d-diverse cluster*, if it contains $d$ disjoint cliques $K_d$ of size $d$. Analogously a graph contains a *d-diverse co-cluster*, if it contains the complement of a $d$-diverse cluster. The following lemma restricts the diversity of clusters and co-clusters in $(H(0,2,\boldsymbol{c}), \overline{H(0,2,\boldsymbol{c})})$-free graphs.

**Lemma 18.** *There is a constant $D = D(\boldsymbol{c}) = \max(2\boldsymbol{c} + 6, 2\boldsymbol{c}^2 + 2)$ such that no $H(0,2,\boldsymbol{c})$ and $\overline{H(0,2,\boldsymbol{c})}$-free graph contains a $D$-diverse cluster and a $D$-diverse co-cluster.*

*Proof.* Suppose $G$ is $(H(0,2,\boldsymbol{c}), \overline{H(0,2,\boldsymbol{c})})$-free, and contains a diverse cluster and diverse co-cluster of diversity $D$. Then $G$ contains a diverse cluster $A$ and a vertex disjoint co-cluster $B$ both of diversity $D/2$. We claim that $B$ contains a co-cluster of diversity $D/2 - 1$ such that every vertex of $B$ is completely adjacent to all but possibly $\boldsymbol{c} - 1$ cliques of $A$: If a vertex $v'$ of $B$ is a neighbor of two vertices contained in different cliques of $A$ then due to the absence of $H(0,2,\boldsymbol{c})$ vertex $v'$ is completely adjacent to all but $\boldsymbol{c} - 1$ cliques of $A$. Suppose there are two vertices in an independent set $I_1$ of $B$ not completely adjacent to all but $\boldsymbol{c} - 1$ cliques and another vertex in a different independent set $I_2$ of $B$ not completely adjacent to all but $\boldsymbol{c} - 1$ cliques of $A$. By the previous statement each of the three vertices has neighbors in at most one clique of $A$. Thus, the three vertices have $D/2 - 3 > \boldsymbol{c}$ cliques in common in which none of the vertices has a neighbor. The three vertices form a $P_2$ and together with one vertex from each of these cliques they form a $H(0,2,\boldsymbol{c})$, a contradiction.

Applying the claim once to $G$ and once to the complement of $G$, we obtain a cluster $A' \leq A$ and a vertex-disjoint co-cluster $B' \leq B$ each of diversity $D/2 - 1$, such that each vertex of $B'$ is completely adjacent to all but $\boldsymbol{c} - 1$ cliques of $A'$, and each vertex of $A'$ is completely not-adjacent to all but $c' - 1$ independent sets of $B'$. We now choose $\boldsymbol{c}$ vertices of $B'$ in different independent sets. Since $D/2 - 1 \geq \boldsymbol{c}^2 > \boldsymbol{c} \cdot (\boldsymbol{c} - 1)$ they have a common neighbor $u$ in $A'$. This is a contradiction, since the vertices in $A'$, and in particular $u$, are completely non-adjacent to all but $\boldsymbol{c} - 1$ vertices in $A'$. □

**Lemma 19.** *Isomorphism of $(H(0, 2, \boldsymbol{c}), \overline{H(0, 2, \boldsymbol{c})})$-free graphs is decidable in polynomial time.*

*Proof.* Let $d = \max(D, 3\boldsymbol{c})$, where $D$ is chosen according to Lemma 18. If the input graphs do not agree on the containment of $d$-diverse (co-)clusters, they are not isomorphic. (Clearly, for any constant $d$, we can find in polynomial time a $d$-diverse (co-)cluster as a subgraph of the input, or determine, that no such subgraph exists.)

By Lemma 18 any $(H(0, 2, \boldsymbol{c}), \overline{H(0, 2, \boldsymbol{c})})$-free graph can only contain a $d$-diverse cluster or a $d$-diverse co-cluster. If the input graphs neither contain a $d$-diverse cluster nor a $d$-diverse co-cluster, we can individualize a bounded number of vertices, such that the evolving color classes are all of bounded degree or bounded co-degree. By Lemma 14 we then can determine isomorphism in polynomial time.

It remains therefore to consider the case that both graphs contain a $d$-diverse cluster but no $d$-diverse co-cluster (w.l.o.g. otherwise taking complements). We find, by guessing, in each input graph a $d$-diverse cluster, such that, if there is an isomorphism between the inputs, there is also an isomorphism that maps the diverse clusters.

We focus on the graph $G_1$. We individualize the vertices according to the $d$-diverse cluster. Consider a color class red. If the red vertices have neighbors in exactly one of the cliques of the cluster, then they are adjacent to all of the vertices in the clique, since otherwise we can find an $H(0, 2, \boldsymbol{c})$ (using single vertices from $\boldsymbol{c}$ other cliques of the cluster for the independent set). Furthermore, the red vertices form a clique. If the red vertices have neighbors in at least two of the cliques, then they are completely adjacent to at least $d - b + 1$ cliques since, again, otherwise we find an $H(0, 2, \boldsymbol{c})$.

The vertices that are adjacent to none of the cliques form a cluster graph $A$. The set of all vertices that are adjacent to more than one clique form a co-cluster $B$, since each triple of such vertices shares $d - 3(b + 1)$ cliques to which all three vertices are completely adjacent. By assumption $B$ is not $d$-diverse, i.e., it contains less than $d$ independent sets of size at least $d$. We can thus individualize $d$ vertices such that in $B$ color classes each of bounded degree or bounded co-degree emerge. For this we guess one vertex in each independent set of size larger than $d$. Clearly there is bounded degree in each of these independent sets (namely degree 0) and we have bounded co-degree among all vertices in the color class containing all independent sets of size less than $d$ (namely co-degree bounded by $d$). If $A$ is not $d$-diverse we can individualize vertices there as well and apply Lemma 14 to decide isomorphism in polynomial time. Thus suppose otherwise. Then $A$ has at least $d$ components of size at least $d$. Vertices which are neighbors to exactly one of the cliques cannot be a neighbor of a vertex in $A$, since otherwise they form an $H(0, 2, \boldsymbol{c})$. We are left with color classes which all have bounded color valence (except possibly $A$) and between the classes there is bounded color valence, except possibly between $A$ and $B$.

Since in each color class of $B$ vertices have degree bounded by $d$ or co-degree bounded by $d$, a vertex from $A$ has bounded degree or bounded co-degree towards any color class in $B$ (by Lemma 14). We show that all except finitely many cliques exclusively have vertices of bounded co-degree towards $B$:

**Claim 1.** *There is a bounded number of cliques in $A$ that contain a vertex that does not have bounded co-degree towards some color class of $B$.*

Assume otherwise and pick such a vertex in each such clique. The chosen vertices must each have bounded degree into some large color class of $B$. Therefore a large subset of them has a common non-neighbor $p \in B$. Together with two neighbors $u$ and $v$ of $p$ in two of the $d$ individualized cliques they form an $H(0, 2, \boldsymbol{c})$. This proves Claim 1.

We color these (bounded many) cliques different and consider the graph $A'$ induced by the remaining vertices. We will now show the following claim:

**Claim 2.** *There is a constant $K$ (independent of the input graph $G_1$ we started with), such that for each clique $C$ of $A'$ there are only at most $K$ types of neighborhoods of vertices outside the clique, i.e., the set $\{N(v) \setminus C \mid v \in C\}$ has size at most $K$.*

It suffices to show the statement for each color-class outside of $A'$, i.e., to show that for each color class $R$ there is a constant $K_R$, such that the set $\{(N(v) \setminus A') \cap R \mid v \in C\}$ has size at most $K_R$. For the color classes consisting of vertices not in $B$ the statement is trivial. For small color classes of $B$ the statement is trivial. For small cliques $C$ the statement is trivial.

First, suppose some large color class $B_i$ of $B$ has bounded co-degree. Further suppose there is a vertex in a clique $C$ of $A'$ not adjacent to a vertex $v \in B$. We can chose a vertex in a different clique of $A$ adjacent to $v$. We now find a fixed number of vertices in $B$ adjacent to both vertices and adjacent to each other and to $v$. This produces a $\overline{H(0, 2, \boldsymbol{c})}$. Thus every vertex of $C$ is a neighbor of every vertex of $B_i$ proving the claim.

Second, suppose some color class $B_j$ of $B$ has bounded degree (in fact such color classes of $B$ are independent sets) and $C$ is a large clique of $A$. We show that there is only a bounded number of vertices in $B_j$ that are not completely adjacent to the clique $C$. Suppose otherwise.

Any vertex $v \in B_j$ not adjacent to all of $C$ is adjacent to a bounded number of vertices in $C$: Any large number of such vertices in $C$ has a common neighbor $v'$ in $B_j$, and therefore some subset of vertices in $C$ would form with $v$ and $v'$ a $\overline{H(0, 2, \boldsymbol{c})}$ otherwise. Suppose now that the number of vertices in $B_j$ that are not completely adjacent to the clique $C$ is large, then there is a vertex $v'$ in the clique that has many non-neighbors in $B_j$. This is a contradiction, since cliques that contain such a $v'$ have been removed in $A'$. This proves Claim 2.

We shrink each clique of $A'$, by removing all but one representative vertex from every neighborhood type. We then color this representative with the color that is the pair of its previous color, and the number of vertices that had the same neighborhood type. This process is reversible and preserves isomorphism. Suppose $A''$ is the shrunken subgraph of $A'$, and $G_1''$ is the shrunken subgraphs of the complete input graph $G_1$.

The whole resulting colored graph $G_1''$ has bounded color valence: Since the cliques in $A'$ are small, the graph induced by $A'$ has bounded degree. Since we had bounded color valence between all color classes except between $A'$ and $B$ in $G_1$ before shrinking the cliques in $A'$, it suffices to consider the adjacencies between $A''$ and $B$. Due to the shrinking process all cliques in $A''$ are small, every vertex in $B$ has a most $c - 1$ cliques in $A'$ to which it is not completely adjacent, and thus has bounded co-degree towards $A''$. Every vertex in $A'$ has bounded co-degree towards every color class of $B$. This holds in particular for vertices in $A''$.

Repeating the coloring and shrinking process with input graph $G_2$ we obtain two graphs $G_1''$ and $G_2''$ which are isomorphic, if and only if $G_1$ are isomorphic $G_2$. Since the new graphs have bounded color valence, we can use Theorem 5 to decide isomorphism. □

## C.4  Case IV: $c \geq 2$, $c' \leq 1$, $b \geq 1$, and $b' \in \{1, 2\}$

Finishing the proof of Theorem 4 we now treat the remaining case $c \geq 2$, $c' \leq 1$, $b \geq 1$, and $b' \in \{1, 2\}$. It suffices to show that the Graph Isomorphism problem is polynomially solvable on $(H(0, b, c), \overline{H(0, 2, 1)})$-free graphs.

**Lemma 20.** *The isomorphism problem on $(H(0, b, c), \overline{H(0, 2, 1)})$-free graphs is solvable in polynomial time.*

*Proof.* It suffices to solve the problem for connected graphs, as the isomorphism check may be performed component-wise. Furthermore, since isomorphism for the $P_4$-free graphs is polynomially solvable (see Theorem 2), it suffices to perform isomorphism tests on graphs that contain a $P_4$. By Lemma 1 the remaining graphs are $K_3$-free.

We now show the statement by induction on $c$. For $c = 0$ the polynomial-time solvability follows from Ramsey's Theorem, since the resulting graphs have bounded degree (as already described in the introduction).

By induction, suppose that both connected input graphs $G_1$ and $G_2$ contain $H(0, b, c-1)$ as a subgraph. By guessing vertices we obtain in each input graph $G_i$ a subgraph $H_i = H(0, b, c-1)$, such that if the input graphs are isomorphic, then there is an isomorphism that maps these graphs. We re-color the vertices according to their adjacency to $H_i$. Since the input graphs are $H(0, b, c)$-free, every vertex is a neighbor of some vertex in $H$. Thus, by $K_3$-free, the arising color classes form independent sets. By Lemma 14 we have bounded color valence and with Theorem 5 we can solve the isomorphism problem in polynomial time. □

## D  Proof of Lemma 11 and Theorem 6

*Proof of Lemma 11.* Since $G$ is split, it is $2K_2$-free and therefore contains at most one component that contains an edge. Furthermore every possible (non-trivial) component of $G$ is a path-star, and does not contain a path $P_5$ of length 4, since $2K_2 \leq P_5$.

If $G$ contains an edge then there are at most 2 vertices that are not connected to the edge: One of the endpoints must be in the clique part of any split decomposition. Since $G$ fulfills split graph condition 2, at most 2 vertices may be non-connected to a vertex in the clique part.

Thus, if $G$ is a non-trivial path with added isolated vertices, then there are most two such vertices. Furthermore, if the path has length 3 there is a most 1 such vertex. Thus if $G$ contains no vertex of degree at least 3, then it is an induced subgraph of one of the graphs $P_4 \dot\cup K_1$, $P_3 \dot\cup I_2$, or it is an independent set.

The graph $G$ does not contain a vertex of degree 4: Since $G$ is a tree the neighbors of a vertex form an independent set. If $v$ is a vertex of degree at least 4, then in any split decomposition $v$ must be in the partition that forms the clique and at least 3 of the neighbors must be within the partition that forms the independent set. Then $G$ would not fulfill split graph condition 1.

Suppose $v$ is a vertex of degree 3. There is at most one induced path of length 2 emanating from $v$. Otherwise $G$ contains two disjoint edges and is therefore not split. Likewise there is no induced path of length 3 emanating from $v$. If there is an induced path of length 2 emanating from $v$, then the neighbor of $v$ along this path must be placed in the partition class that forms a clique in any split decomposition. This vertex has two non-neighbors within the connected component of $v$, thus there is no isolated vertex in $G$. This leaves $K_{1,3} \dot\cup I_2$ and the fork as the two maximal graphs with a vertex of degree 3. □

*Proof of Theorem 6.* Due to Lemma 2 we may assume that one of the graphs $H_1$ and $H_2$ is a union of path-stars and one is a complement of a union of path-stars (or else we have GI-completeness on $(H_1, H_2)$-free). W.l.o.g. $H_1$ is union of path-stars, since fulfillment of all split conditions is unaffected by taking complements (e.g., condition 1 for a graph $H$ implies condition 4 for its complement).

Since $H_1$ is non-basic, by Lemma 11, $H_1$ is **(a)** an induced subgraph of $P_4$ or one of the graphs **(b)** $P_4 \mathbin{\dot\cup} K_1$, **(c)** $K_2 \mathbin{\dot\cup} I_2$, **(d)** $P_3 \mathbin{\dot\cup} I_2$, **(e)** $K_{1,3}$, **(f)** $K_{1,3} \mathbin{\dot\cup} K_1$, **(g)** $K_{1,3} \mathbin{\dot\cup} I_2$, or **(h)** the fork. We distinguish the different cases of $H_1$.

**(a) $H_1 \leq P_4$:** By Theorem 2 Graph Isomorphism is polynomial-time solvable on $H$-free graphs when $H$ is an induced subgraph of $P_4$.

If $H_1$ is a union of path-stars and not a subgraph of the $P_4$, then unless $H_2$ is the complement of a union of path-stars, by Lemma 2 and 3, Graph Isomorphism is GI-complete on $(H_1, H_2)$-free. We thus assume that $H_2$ is the complement of a union of path-stars.

We make frequent use of GI-completeness for $(diamond, K_{1,3}, C_4)$-free graphs and its complement class $(K_2 \mathbin{\dot\cup} I_2, \overline{K_{1,3}}, 2K_2)$-free (Lemma 9). We also use the following claim.

**Claim 3.** *If $H$ is a non-basic union of path-stars with $H \not\leq P_4$ then $K_2 \mathbin{\dot\cup} I_2 \leq H$ or $2K_2 \leq H$ or $H$ is a subgraph of $K_{1,t} \mathbin{\dot\cup} K_1$ for some integer $t \geq 3$.*

*Proof.* The non-basic graph $H$ contains at least one edge. Hence it has at most 2 connected components, or else $K_2 \mathbin{\dot\cup} I_2 \leq H$ follows. If it has more than one nontrivial component then $2K_2 \leq H$. If $H$ has a vertex $v$ of degree larger than 2 then either it is a subgraph of $K_{1,t} \mathbin{\dot\cup} K_1$ for some integer $t$ or some edge is not incident with $v$, in which case $H$ contains $K_2 \mathbin{\dot\cup} I_2$. Suppose now $H$ does not have a vertex of degree larger than 2, which implies $H$ is a union of paths. If $H$ is connected it must be a path $P_i$ on at least 5 vertices, since $H \not\leq P_4$. But then $2K_2 \leq P_5 \leq H$. If $H$ is disconnected, it is a path with an additional isolated vertex, i.e., $H = P_i \mathbin{\dot\cup} K_1$. Since $P_2 \mathbin{\dot\cup} K_1 \leq P_4$ and $P_3 \mathbin{\dot\cup} K_1$ is basic, we conclude that $i \geq 4$. This implies however that $K_2 \mathbin{\dot\cup} I_2 \leq P_4 \mathbin{\dot\cup} K_1 \leq H$. □

**(e,f,g,h) $H_1 \in \{K_{1,3}, K_{1,3} \mathbin{\dot\cup} K_1, K_{1,3} \mathbin{\dot\cup} I_2, \text{fork}\}$:** For all choices of $H_1$ we have $K_{1,3} \in H_1$. Since we assume $\overline{H_2} \not\leq P_4$ by the claim there are three options for $\overline{H_2}$. In the first two, when $K_2 \mathbin{\dot\cup} I_2 \leq \overline{H_2}$ or $2K_2 \leq \overline{H_2}$ we conclude GI-completeness with Lemma 9. Otherwise $\overline{H_2} \leq K_{1,t} \mathbin{\dot\cup} K_1$ with $t \geq 3$. If $H_1$ is the fork (case h), then GI-completeness follows from Lemma 9 since $2K_2 \mathbin{\dot\cup} I_2 \leq H_1$ and $K_{1,3} \leq \overline{H_2}$. If $H_1$ is not the fork (case e–g) Theorem 4 applies.

**(b) $H_1 = P_4 \mathbin{\dot\cup} K_1$:** Since $H_2$ is a co-forest, if $H_2$ has at least 7 vertices, then it is easy to see that it contains $K_4$ as induced subgraph. By Theorem 3 Graph Isomorphism is GI-complete for the class $(P_4 \mathbin{\dot\cup} K_1, K_4)$-free $\subseteq (H_1, H_2)$-free. Otherwise, $\overline{H_2}$ is a disjoint union of path-stars on at most 6 vertices. If it contains a vertex of degree at least 3, and hence a claw $K_{1,3}$, then GI-completeness of $(H_1, H_2)$-free follows from the class $(K_2 \mathbin{\dot\cup} I_2, \overline{K_{1,3}})$-free. In the remaining cases, $\overline{H_2}$ is a nonbasic disjoint union of paths with at most 6 vertices and independent set size at most 3 (hence at most 3 components). As $\overline{H_1}$ is not bipartite, GI-completeness follows also when $\overline{H_2}$ contains $3K_2$ (Lemma 5). Thus, skipping subgraphs of $P_4$ and basic graphs, we have: $\overline{H_2} \in \{P_6, P_5, P_4 \mathbin{\dot\cup} K_1, P_4 \mathbin{\dot\cup} K_2, P_3 \mathbin{\dot\cup} K_2, 2K_2, 2K_2 \mathbin{\dot\cup} K_1, K_2 \mathbin{\dot\cup} I_2\}$.

**(c,d) $H_1 \in \{K_2 \mathbin{\dot\cup} I_2, P_3 \mathbin{\dot\cup} I_2\}$:** We have $K_2 \mathbin{\dot\cup} I_2 \leq H_1$, thus if $\overline{K_{1,3}} \leq H_2$ then GI-completeness follows from $(K_2 \mathbin{\dot\cup} I_2, \overline{K_{1,3}})$-free $\subseteq (H_1, H_2)$-free (Lemma 9). Thus it remains to consider $\overline{K_{1,3}} \not\leq H_2$ or, equivalently, $K_{1,3} \not\leq \overline{H_2}$, restricting the path-star $\overline{H_2}$ to a disjoint union of paths (as discussed above). Furthermore, since the complement of $H_1$ is not bipartite (as $K_3 \leq \overline{H_1}$), it follows that $\overline{H_2}$ does not contain $3K_2$, $2K_2 \cup I_2$, and $P_4 \mathbin{\dot\cup} I_2$ or GI is GI-complete on $(H_1, H_2)$-free (Lemma 5). From $3K_2 \not\leq \overline{H_2}$ we get that there are at most two nontrivial components (i.e., paths). Since $\overline{H_2}$ is nonbasic, is has at least one nontrivial component. Thus $\overline{H_2} = P_t \mathbin{\dot{\cup}} P_{t'} \mathbin{\dot{\cup}} I_{t''}$ for integers $t$, $t'$, and $t''$ with $t \geq 2$.

Concretely, we get the following choices, ignoring subgraphs of $P_4$ and basic graphs. Suppose first that $\overline{H_2}$ contains one path.

- If $\overline{H_2}$ contains at least two isolated vertices, then the path has at most 3 vertices, else $P_4 \mathbin{\dot{\cup}} I_2 \leq \overline{H_2}$. All ensuing cases are tractable according to Part 3 of Theorem 4.

- If there is one isolated vertex, then the path has at most 5 vertices, else $P_4 \mathbin{\dot{\cup}} I_2 \leq P_6 \mathbin{\dot{\cup}} K_1 \leq \overline{H_2}$. We get $\overline{H_2} \in \{P_5 \mathbin{\dot{\cup}} K_1, P_4 \mathbin{\dot{\cup}} K_1\}$.

- If there is no isolated vertex, then the path has at most 7 vertices, else $P_4 \mathbin{\dot{\cup}} I_2 \leq P_8 \leq \overline{H_2}$. We get $\overline{H_2} \in \{P_7, P_6, P_5\}$.

If $\overline{H_2}$ contains two paths, then it has at most one isolated vertex as $2K_2 \mathbin{\dot{\cup}} I_2 \not\leq \overline{H_2}$:

- If there is one isolated vertex, then none of the paths contains 4 or more vertices, as $P_4 \mathbin{\dot{\cup}} I_2 \not\leq \overline{H_2}$. We get $\overline{H_2} \in \{P_3 \mathbin{\dot{\cup}} P_3 \mathbin{\dot{\cup}} K_1, P_3 \mathbin{\dot{\cup}} K_2 \mathbin{\dot{\cup}} K_1, 2K_2 \mathbin{\dot{\cup}} K_1\}$.

- If there is no isolated vertex, then none of the paths contains 5 or more vertices, as $3K_2 \not\leq \overline{H_2}$. The cases $\overline{H_2} \in \{P_4 \mathbin{\dot{\cup}} P_4, P_4 \mathbin{\dot{\cup}} P_3\}$ are excluded as $P_4 \mathbin{\dot{\cup}} I_2 \not\leq \overline{H_2}$. We get $\overline{H_2} \in \{P_4 \mathbin{\dot{\cup}} K_2, P_3 \mathbin{\dot{\cup}} P_3, P_3 \mathbin{\dot{\cup}} K_2, 2K_2\}$. □